# Global-mean surface air temperature change overestimates global warming rate

**Jeremy Cheuk-Hin Leung[1,2], Qiuying Gan[3], Banglin Zhang[1,2,4,5]***

**Affiliations**

[1] College of Meteorology and Oceanography, National University of Defense Technology, Changsha, China

[2] Laboratory of Atmospheric Environmental Monitoring and Early Warning for Low-Altitude Economy, National University of Defense Technology, Changsha, China

[3] School of Atmospheric Sciences and Key Laboratory of Tropical Atmosphere-Ocean System Ministry of Education Sun Yat-Sen University, Southern Marine Science and Engineering Guangdong Laboratory (Zhuhai), Zhuhai, China

[4] College of Atmospheric Sciences, Lanzhou University, Lanzhou, China

[5] Key Laboratory of High Impact Weather (Special), China Meteorological Administration, Changsha, China

*** Corresponding author:** Banglin Zhang (E-mail: zhangbanglin24@nudt.edu.cn)

**Abstract**

Current climate policies are targeted at slowing down the global warming rate, or the global mean surface air temperature (SAT) change ($\Delta T_{mean}$), which is measured by the arithmetic mean approach under the assumption that SAT changes are symmetrically distributed. However, in reality, the SAT change is asymmetric in nature and its influence on the $\Delta T_{mean}$ interpretation seldom received attention in previous research about climate change. This study theorizes, based on the image histogram approach, that while $\Delta T_{mean}$ measures the Earth's overall SAT change, it yields a value larger than the global-scale SAT change ($\Delta T_{gs}$) because of the asymmetrical distribution of SAT change. Results show that $\Delta T_{mean}$ is greater than that based on $\Delta T_{gs}$ by 0.19–0.23°C from 2000–2019, relative to the global SAT in 1979. In future climate projections, where more significant inhomogeneous warming is expected, the disagreement between $\Delta T_{mean}$ and $\Delta T_{gs}$ reach 0.27–0.54°C by the end of the 21st century (2080–2099) under different emission scenarios. The large difference between $\Delta T_{mean}$ and $\Delta T_{gs}$ implies that there is a net positive regional-scale warming effect over the globe which is mainly contributed by Arctic Amplification. This paper constitutes a warning that extreme regional warming effects could have large impacts on the interpretation of $\Delta T_{mean}$, which is often considered in climate assessments and policies making, and illustrates the limitations and cautions inherent in using $\Delta T_{mean}$ as the only indicator of the global warming rate.

## 1. Introduction

Climatologists have been paying close attention to the impacts of global warming on the environment, society and economy (Burke et al. 2018; Hughes et al. 2018; Patricola and Wehner 2018; Reich et al. 2018; Williams et al. 2020; Laufkötter et al. 2020; Cohen et al. 2020; Woolway et al. 2021) in both the past and future. Today, the global warming rate, measured by global mean surface temperature (SAT) change, is the most commonly used indicator of climate change mitigation by scientists and policymakers (Solomon et al. 2007; Meinshausen et al. 2009; Hansen et al. 2010; Stocker et al. 2013; Tebaldi and Arblaster 2014; Burke et al. 2015; Masson-Delmotte et al. 2021). In 2015, scientists and policymakers reached a strong consensus to limit the global mean SAT increase to well below 2°C, and preferably to 1.5°C, above the preindustrial level to achieve sustainable development (Rogelj et al. 2015; Hof et al. 2017; Sanderson et al. 2017; Masson-Delmotte et al. 2018; Hoegh-Guldberg et al. 2019; Duan et al. 2021).

Global mean SAT change ($\Delta T_{mean}$) is a measure of the central value of global SAT change by the arithmetic mean approach, or more specifically the area-weighted mean SAT change. The arithmetic mean approach statistically neglects spatial variations in SAT change and gives an overall estimate of the change of the SAT on Earth, or global-scale SAT change. Because of its simplicity, $\Delta T_{mean}$ has been the only measure of the global warming rate. However, the arithmetic mean approach is known to have certain limitations; it gives the best measure of the central value only when sample data are symmetrically distributed, and arithmetic mean values are easily influenced by outliers, missing data, and asymmetrical data distribution etc. (Hays 1994; Walpole et al. 2016). A common example is global wealth distribution, in which the richest 1% of people account for more than 40% of the world's wealth, while the poorest 50% of the population accounts for only approximately 1.5% of global wealth (Shorrocks et al. 2020). In such a highly skewed wealth distribution, the wealth level of the majority cannot be estimated by the simple arithmetic mean approach and hence other measures, such as median, are also given as a reference.

In reality, the global SAT and its changing rate are also asymmetric in nature given the climatological land-sea and meridional variation and the spatially inhomogeneous warming effect (Gan et al. 2026a). Previous studies have reported that the rate of SAT change varies in different regions; i.e., some places warm more quickly, while others warm slowly or even experience cooling trends (Serreze and Francis 2006; Hansen et al. 2010; Huang et al. 2012; Meehl et al. 2012; Stocker et al. 2013; Ji et al. 2014; Walsh et al. 2014; Qian et al. 2016, 2019). Hence, despite its simplicity, $\Delta T_{mean}$ may not be an optimal measure of the central value of global SAT change because of the limitation of arithmetic mean approach on asymmetrical sample data. In other words, the asymmetrically distributed SAT and SAT change may affect our interpretation of $\Delta T_{mean}$, which has been considered as an indicator of global-scale temperature change.

While most studies have treated $\Delta T_{mean}$ as the only measure of the global warming rate, its limitations and downside were rarely discussed. Particularly, the asymmetrical nature of SAT change and its impacts on the interpretation of $\Delta T_{mean}$ seldom received attention in previous studies on climate change. The global warming rate is one of the most fundamental indicators of climate change, and climate policies are targeted at controlling the global warming rate. An accurate interpretation of $\Delta T_{mean}$ could have a decisive impact on the strategies and policies for mitigating climate change as well as improve our understanding of the mechanism and impacts of climate change. Thus, this paper aims at reporting the asymmetrical nature of global SAT and SAT change, examining the relationship between $\Delta T_{mean}$ and global-scale SAT change, and discussing how this could affect our interpretation of the current measure of global warming rate, based on the image histogram approach (Section 2.2).

## 2. Data and methods

### *2.1 Data*

Monthly SAT (2-m air temperature) data from four reanalysis datasets were analyzed in this study. The datasets include (1) the Fifth generation of the European Centre for Medium-Range Weather Forecasts (ECMWF) atmospheric reanalysis (ERA5) (Hersbach et al. 2020) dataset, which has a horizontal resolution of 0.25°×0.25°; (2) the Japanese Meteorological Agency (JMA) 55-year reanalysis (JRA55) (Kobayashi et al. 2015) dataset, with a horizontal resolution of TL319 (approximately 55 km or 0.5625°); (3) the National Centers for Environmental Prediction (NCEP)/National Center for Atmospheric Research (NCAR) Reanalysis I (NCEP1) (Kalnay et al. 1996) dataset, with a 2.5°×2.5° horizontal resolution; and (4) the NCEP/Department of Energy (DOE) Reanalysis II (NCEP2) (Kanamitsu et al. 2002) dataset, with a 2.5°×2.5° horizontal resolution. Annual-mean temperature values were obtained from the averages of monthly values.

Twenty Coupled Model Intercomparison Project (Phase 6) (CMIP6) global climate model simulations of historical runs and future projections under the shared socioeconomic pathway (SSP) 1-2.6, SSP2-4.5, and SSP5-8.5 scenarios (Eyring et al. 2016) were used to assess the future climate projections. The twenty CMIP6 models include ACCESS-CM2, ACCESS-ESM1–5, AWI-CM-1–1-MR, BCC-CSM2-MR, CAMS-CSM1–0, CESM2-WACCM, CIESM, CMCC-CM2-SR5, FGOALS-f3-L, FIO-ESM-2–0, INM-CM4–8, INM-CM5–0, IPSL-CM6A-LR, KACE-1–0-G, KIOST-ESM, MPI-ESM1–2-HR, MPI-ESM1–2-LR, MRI-ESM2–0, NESM3, and NorESM2-LM. The ensemble results in the following analyses are derived form the averages of the twenty CMIP6 models.

### *2.2 Method: Image histogram approach*

The image histogram (hereafter histogram, unless specified) approach helps us explore the asymmetry of global SAT and SAT changes, and how the global SAT evolves in a more comprehensive way. This fundamental statistical method is widely used in image processing and pattern recognition (Motoyoshi et al. 2007; Burger and Burge 2008; Tsurusaki et al. 2014; Kravchenko et al. 2019); it analyzes the probability frequency distribution (PFD) of a numerical variable (e.g., SAT value) to determine the probability at which (or number of pixels in which) particular values appear in an image. The image histogram approach has been applied to climate science research recently, by treating climate data as climate snapshots (Leung et al. 2022; Gan et al. 2023, 2026b; Liu et al. 2025).

In the context of climate analysis, an image histogram displays a PFD of SAT (or SAT change) by measuring the number of grid points (or area) at which particular values appear in a year (or period) (Leung et al. 2022; Liu et al. 2025). PFDs are commonly used in climate research but are usually obtained for a temporal dimension (Perron and Sura 2012; Pendergrass and Gerber 2016; O'Gorman et al. 2018; Tamarin-Brodsky and Hadas 2019), and they indicate how often certain values occur at a given location or in a certain region during a study period. In contrast, the application of an image histogram approach to climate variables is equivalent to a PFD calculated based on the spatial dimension, reflecting how large an area a certain value covers at a certain time. The image histogram approach, similar to other dimensionality reduction methods, has advantages in extracting information that is difficult to observe in a two-dimensional image. In the following analyses, we transform the global annual-mean SAT data into a set of image histograms; then, extract information about the rate of global warming from the changes in histogram properties. The detailed procedure for deriving an SAT histogram is as follows:

(1) For each year, estimate the frequency of each histogram bin by calculating the total area of grid points with SAT values in the range of each bin. The bin width is set to 0.01°C in this study;
(2) Plot the estimated frequency value in the form of a histogram, where the x-axis and y-axis indicate the SAT values and area, respectively;
(3) Depending on the ratio of the horizontal resolution of the SAT data to the bin width, some bins of the plotted histogram may be empty. This issue can be solved by adjusting the histogram bin width or applying a smoothing filter. It is also noted that the total summed frequency of each histogram, which represents the total area, should be constrained to be equal if the study domain does not change over time.

In the following context, different measures of central value of global temperature change are analyzed based on the calculated histograms. The central value, or central tendency, is the number that represents the central location of a dataset or a distribution. The three common measures of central tendency are the mean, median, and mode. The mean, or the arithmetic mean, of a dataset is calculated by simply summing all the values and dividing by the total number of values. The median is the middle value (50th percentile) in a dataset. The mode corresponds to the peak of a distribution or histogram and the SAT value that occurs at the highest number of grid points in the case of the SAT image histogram. The most commonly used measure of the global warming rate, $\Delta T_{mean}$, is equal to the arithmetic mean value of the global SAT change.

Apart from the above three measures, we also examine the global-scale warming rate ($\Delta T_{gs}$), which is defined as the SAT change over the majority of the Earth's surface. Based on the properties of the image histogram, if the Earth's surface uniformly warms by a certain degree, all grid points on Earth warm by the same degree, and the SAT histogram will shift to the right without changing its shape. In this sense, changes in the SAT histogram mean, median, and mode are exactly the same. Given that the SAT histogram shape does not exhibit large variations over time (Fig. 1), a histogram shift could well represent the SAT change experienced at most grid points, which could be referred to as global-scale warming. Thus, one could estimate $\Delta T_{gs}$ by measuring how far the SAT histogram shifts. Technically, $\Delta T_{gs}$ is defined as the horizontal distance (unit: °C) that the SAT histogram shifts to the right (left). In this study, shifts are identified based on the maximum cross-correlation between the SAT histogram in each year and that for the base year of 1979. Namely, we define $\Delta T_{gs}$ as the degree (or horizontal distance, unit: °C) to which the SAT histogram in a given year shifts to the right considering the similarity with the SAT histogram for 1979. $\Delta T_{gs}$ estimates the SAT change that likely occurs over the largest proportion of grid points and is insensitive to extreme values compared to the arithmetic mean approach. Estimating the similarity of SAT histogram by root-mean-square difference also gives the same result as that estimated by correlation coefficient.

The statistical significance of all linear trend analyses performed in this study was tested by a two-tailed Student's t-test.

## 3. Results

### *3.1 The asymmetrical nature of global SAT and SAT change*

Global warming has been observed in industrial years, but global SAT observations at a high spatial resolution have only been available for the past four decades, or since the satellite era began. Based on the ERA5 reanalysis, state-of-the-art reanalysis data (Hersbach et al. 2020), the observed global SAT primarily displays an asymmetrical distribution. A distribution or histogram that is not symmetrical is said to be asymmetrical or skewed. The asymmetry of a distribution can be measured by the sample skewness of a dataset, defined as $skewness = \frac{\frac{1}{N}\sum_N (x-\bar{x})^3}{(\frac{1}{N}\sum_N (x-\bar{x})^2)^{3/2}}$, where $x$ represents the sample data, $\bar{x}$ is the mean value of $x$ and $N$ is the total number of samples. An asymmetrical distribution has a non-zero skewness, and may have different mean, median and mode values. Positive skewness represents a positively skewed histogram, with a long tail on the right of the distribution, and vice versa. The skewness of global SAT ranges from -1.60 to -1.44 from 1979–2020, which implies that observed SAT is a negatively skewed distribution. The SAT values are more concentrated in the right portion (high SAT) of the histogram and there is a long tail on the left side (low SAT) of the histogram. The entire SAT distribution shifted slightly to the right with very small changes in shape from the 1980s to the 2010s (Fig. 1). Because of these slight changes in shape, the mean, median, and mode values of the SAT distribution increase at different rates.

The shape change of the SAT distribution indicates not only that the SAT itself has a skewed distribution but also that SAT change is asymmetrically distributed. As an example, we investigate the two-decade SAT change between 1980–1999 and 2000–2019. Again, the SAT histogram shifts to the right, with a slight change in shape. The $\Delta T_{mean}$ rises by 0.39°C, from 13.94°C to 14.33°C, and the changes in the median ($\Delta T_{median}$) and mode ($\Delta T_{mode}$) SATs are comparatively smaller (Figs. 1a and 1b). These differences are directly associated with the asymmetrical distribution of SAT change, which exhibits a positive skew and a long tail on the right of histogram (Fig. 1c), implying the existence of extremely large positive values of SAT change that could influence the extent of $\Delta T_{mean}$. From 1980–1999 to 2000–2019, the maximum SAT change ($\Delta T_{max}$) value reaches 4.14°C (3.75°C deviated from $\Delta T_{mean}$) but the minimum value ($\Delta T_{min}$, -1.08°C) is only 1.47°C different from the mean value, indicating a rather large spatial variance of SAT change over the world. Because of the asymmetrical nature of SAT change, $\Delta T_{mean}$ between the two periods exceeds the median ($\Delta T_{median}$, 0.30°C) and mode ($\Delta T_{mode}$, 0.26°C) SAT changes by 0.09°C and 0.13°C, respectively.

The term global warming is generally referred to as the Earth's overall or global-scale temperature change. Hence, we here particularly focus on examining the relationship between $\Delta T_{mean}$ and $\Delta T_{gs}$, i.e., the degree of global-scale warming (see Section 2.2). Comparison results show that from 1980–1999 to 2000–2019 $\Delta T_{mean}$ (0.39°C) is 0.11°C greater than $\Delta T_{gs}$ (0.28°C) (Fig. 1c). The comparatively large $\Delta T_{mean}$ suggests that while the global mean SAT change measures the overall temperature over the globe, it yields an overestimated rate of global-scale warming and there is a net positive warming effect of regional-scale temperature change. In other words, regional-scale warming is greater than regional-scale cooling in recent decades and their net effect also contributes to the global mean SAT rise, other than the global-scale temperature increase. The contributions of global-scale temperature change and regional-scale temperature change to $\Delta T_{mean}$ from 1980–1999 to 2000–2019 are 71.8% and 28.2%, respectively.

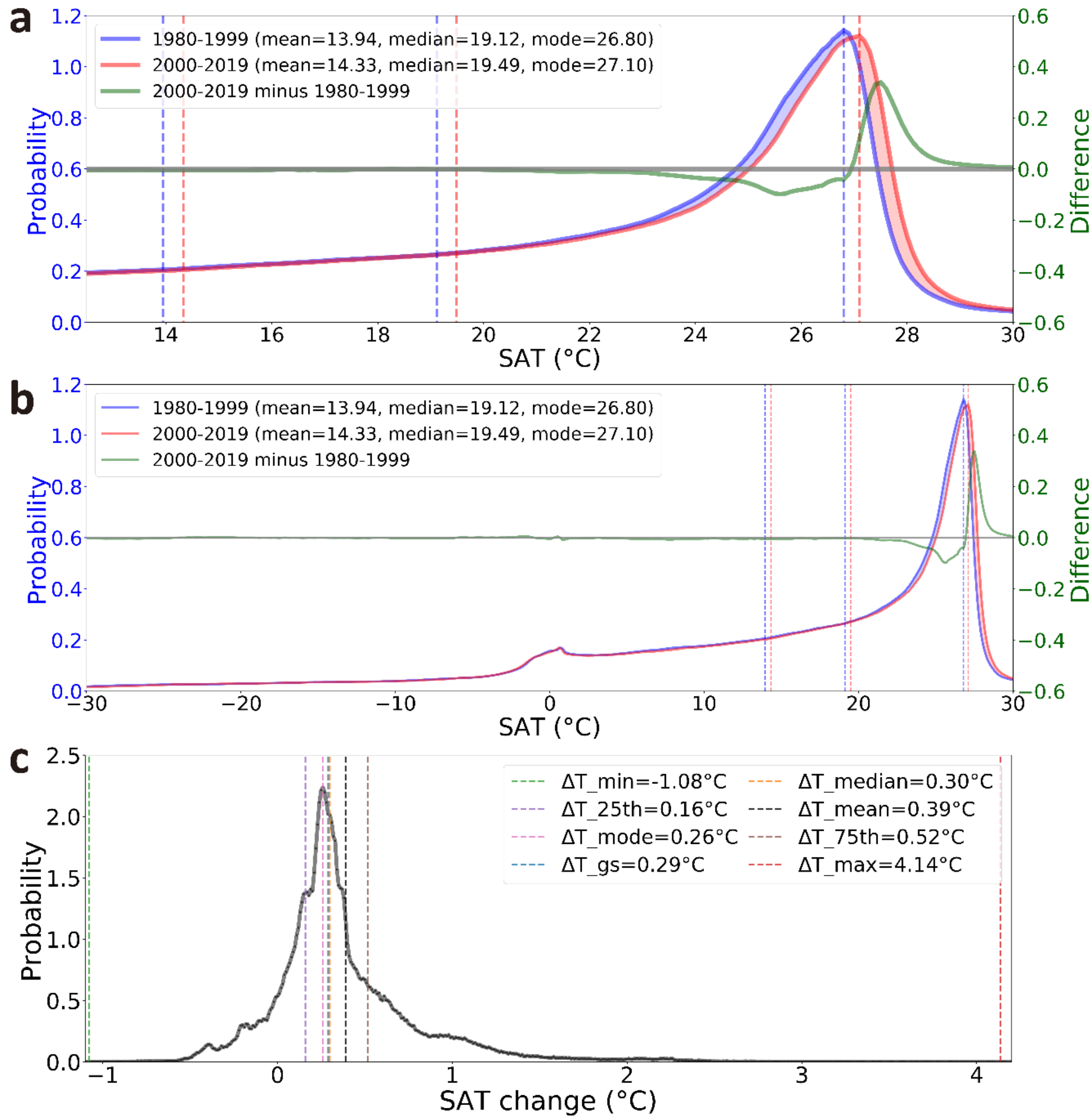

**Figure 1.** The asymmetrical distribution of the SAT and SAT change. (**a**) Two-decadal mean SAT histograms from 1980–1999 (blue) and 2000–2019 (red) and their difference (2000–2019 minus 1980–1999, green). (**b**) Same as (**a**), except that the x-axis ranges from -30°C to 30°C. The shaded region between the red and blue lines in (**a**) and (**b**) indicates increases and decreases in the SAT probability (i.e., area), respectively. The values of the mean (unit: °C), median (unit: °C), and mode (or histogram peak, unit: °C) SAT are denoted in the figure legends and are indicated by dashed lines. (**c**) Histogram of the two-decadal SAT change from 1980–1999 and 2000–2019, in which dashed lines indicate different metrics (mean, median, mode, minimum, and maximum values, and the $10^{th}$, $25^{th}$, $75^{th}$ and $90^{th}$ percentile) of SAT change (unit: °C) and $\Delta T_{gs}$ (unit: °C), as marked in the figure legend. The probability in (**a**)–(**c**) indicates the proportion of the Earth surface area where a certain value of SAT or SAT change is observed (see Section 2). The results are based on ERA5 SAT data.

### *3.2 Impacts of asymmetrical SAT change on $\Delta T_{mean}$ in recent decades*

We further examine the differences between $\Delta T_{mean}$ and $\Delta T_{gs}$ and other metrics of each year's global SAT change from 1979 to 2019. Time series analyses on annual mean SAT show that although $\Delta T_{median}$, $\Delta T_{mode}$ and $\Delta T_{gs}$ exhibit highly correlated variability with $\Delta T_{mean}$, the series of $\Delta T_{median}$, $\Delta T_{mode}$ and $\Delta T_{gs}$ are generally at a level that is lower than $\Delta T_{mean}$. The increasing trend of $\Delta T_{mean}$ (0.19°C/decade), consistent with previously reported (Masson-Delmotte et al. 2018), is greater than those of $\Delta T_{median}$, $\Delta T_{mode}$ and $\Delta T_{gs}$ respectively by 0.03, 0.05 and 0.04°C/decade during the 41-year study period (Fig. 2a). This suggests that the trends of $\Delta T_{median}$ and $\Delta T_{mode}$, compared to $\Delta T_{mean}$, are more consistent with the global-scale SAT change. Relative to the SAT in the base year of 1979, $\Delta T_{mean}$ (0.52°C) from 2000–2019 was 0.19°C larger than $\Delta T_{gs}$ (0.33°C), indicating a big gap between $\Delta T_{mean}$ and the global-scale warming rate (Table 1). The relatively high global warming rate through the arithmetic mean approach is also evident in other reanalysis datasets, including the JRA55, NCEP1, and NCEP2 datasets, which give values of $\Delta T_{mean}$ from 2000–2019 greater than $\Delta T_{gs}$ by 0.20–0.23°C relative to the SAT in 1979 and show larger $\Delta T_{mean}$ trends than the others (Table 1 and Fig. S1).

**Table 1.** $\Delta T_{mean}$ and $\Delta T_{gs}$ and their differences (unit: °C) from 2000–2019, 1995–2014, and 2000–2019 relative to the SAT in the base year of 1979. The results are obtained from reanalysis data (ERA5, JRA55, NCEP1, and NCEP2) and the ensemble results of CMIP6 model simulations (historical run, SSP1-2.6, SSP2-4.5, and SSP5-8.5). Values given in parentheses are the differences in percentages.

| SAT change | ERA5 | JRA55 | NCEP1 | NCEP2 | Historical run | SSP1-2.6 | SSP2-4.5 | SSP5-8.5 |
|---|---|---|---|---|---|---|---|---|
| $\Delta T_{mean}$ (2000–2019) | 0.52 | 0.42 | 0.47 | 0.47 | - | - | - | - |
| $\Delta T_{gs}$ (2000–2019) | 0.33 | 0.22 | 0.24 | 0.25 | - | - | - | - |
| $\Delta T_{mean}$ - $\Delta T_{gs}$ (2000–2019) | 0.19 (57.6%) | 0.20 (90.9%) | 0.23 (95.8%) | 0.22 (88.0%) | - | - | - | - |
| $\Delta T_{mean}$ (1995–2014) | 0.39 | 0.33 | 0.33 | 0.34 | 0.52 | - | - | - |
| $\Delta T_{gs}$ (1995–2014) | 0.24 | 0.16 | 0.17 | 0.18 | 0.42 | - | - | - |
| $\Delta T_{mean}$ - $\Delta T_{gs}$ (1995–2014) | 0.15 (62.5%) | 0.17 (106.3%) | 0.16 (94.1%) | 0.16 (88.9%) | 0.10 (23.8%) | - | - | - |
| $\Delta T_{mean}$ (2080–2099) | - | - | - | - | - | 1.58 | 2.49 | 4.27 |
| $\Delta T_{gs}$ (2080–2099) | - | - | - | - | - | 1.31 | 2.14 | 3.73 |
| $\Delta T_{mean}$ - $\Delta T_{gs}$ (2080–2099) | - | - | - | - | - | 0.27 (20.6%) | 0.35 (16.4%) | 0.54 (13.4%) |

The disagreement among different measures of the global SAT change illustrates the sensitivity of $\Delta T_{mean}$ calculation to the asymmetrical nature of SAT change. Specifically, extremely large values in the positively skewed SAT change distribution, which are the outcome of regional or local warming effects, could lead to a larger global mean SAT increase compared to the global-scale SAT rise. Consequently, $\Delta T_{mean}$, which has been used to present the degree of global-scale warming, includes and is partly contributed by the net effect of regional-scale warming. This is clearly shown from the comparatively slower trends of $\Delta T_{median}$, $\Delta T_{mode}$ and $\Delta T_{gs}$, which are mathematically less sensitive to extreme large (or small) values in calculation (Fig. 2a).

The inhomogeneous spatial distribution of surface warming has already been widely reported in numerous research, as demonstrated by the exceptional SAT increase over the Arctic region and the relatively slow warming trend in the Southern Hemisphere (Solomon et al. 2007; Meinshausen et al. 2009; Hansen et al. 2010; Stocker et al. 2013; Tebaldi and Arblaster 2014; Burke et al. 2015) (Fig. S2). The observed inhomogeneous warming in the past four decades is positively skewed overall, resulting in changes in the shape of the SAT histogram and enlarging the difference between $\Delta T_{mean}$ and $\Delta T_{gs}$ (Fig. 1). In Fig. 2a, the SAT change values that are more frequently observed over the world in each year (marked by red and orange shadings) are tightly linked with the $\Delta T_{median}$, $\Delta T_{mode}$ and $\Delta T_{gs}$ series, but the $\Delta T_{mean}$ series is obviously above these values. The faster $\Delta T_{mean}$ increasing rate is a result of the more positively skewed SAT change distribution in recent years, with more large SAT change values observed, as noted by the much faster increasing trend of $\Delta T_{max}$ (1.20 °C/decade) compared to the others (Fig. 2b). We find that while the variance in the SAT distribution decreased by 1.1–2.1%/decade, the SAT histogram skewness increased by 0.3–1.3%/decade (i.e., less negative) from 1979–2019, which indicates that the global SAT distribution is slightly less asymmetrical and less spread out than it was in the past (Fig. S3). The less negatively skewed SAT histogram implies that the extreme large SAT change values come from the cold regions, such as the Arctic region (Fig. S2). These evidence all suggest that extreme regional warming effects could enhance the difference between $\Delta T_{mean}$ and global-scale warming rate.

In particular, our results show that Arctic amplification (Serreze and Francis 2006; Screen and Simmonds 2010; Serreze and Barry 2011; Cohen et al. 2014; Dai et al. 2019) is a main factor that leads to the disagreement between $\Delta T_{mean}$ and $\Delta T_{gs}$. If the SAT changes north of 65°N are excluded, the time series of $\Delta T_{mean}$ move closer to those of $\Delta T_{median}$, $\Delta T_{mode}$ and $\Delta T_{gs}$, with the linear trend differences reduced to only 0.01–0.02°C/decade (Fig. 2c). The SAT change histogram becomes less positively skewed and the $\Delta T_{max}$ series shows a smaller increasing trend (0.52°C/decade) (Fig. 2d), meaning that extreme large values of SAT change mainly come from north of 65°N. These findings suggest that exceptional regional-scale warming over the Arctic region plays a key role in $\Delta T_{mean}$, resulting in an increasing rate faster than that of global-scale warming. The above results show, from past observations, that the calculation of $\Delta T_{mean}$ also includes regional-scale warming effects in addition to those on the global scale, and is particularly sensitive to extreme SAT change values. Based on the linear trend analysis, the global-scale warming trend contributes to 78.9% of the $\Delta T_{mean}$ trend from 1979–2020, and the rest is contributed by the net warming effect of reigonal-scale SAT change.

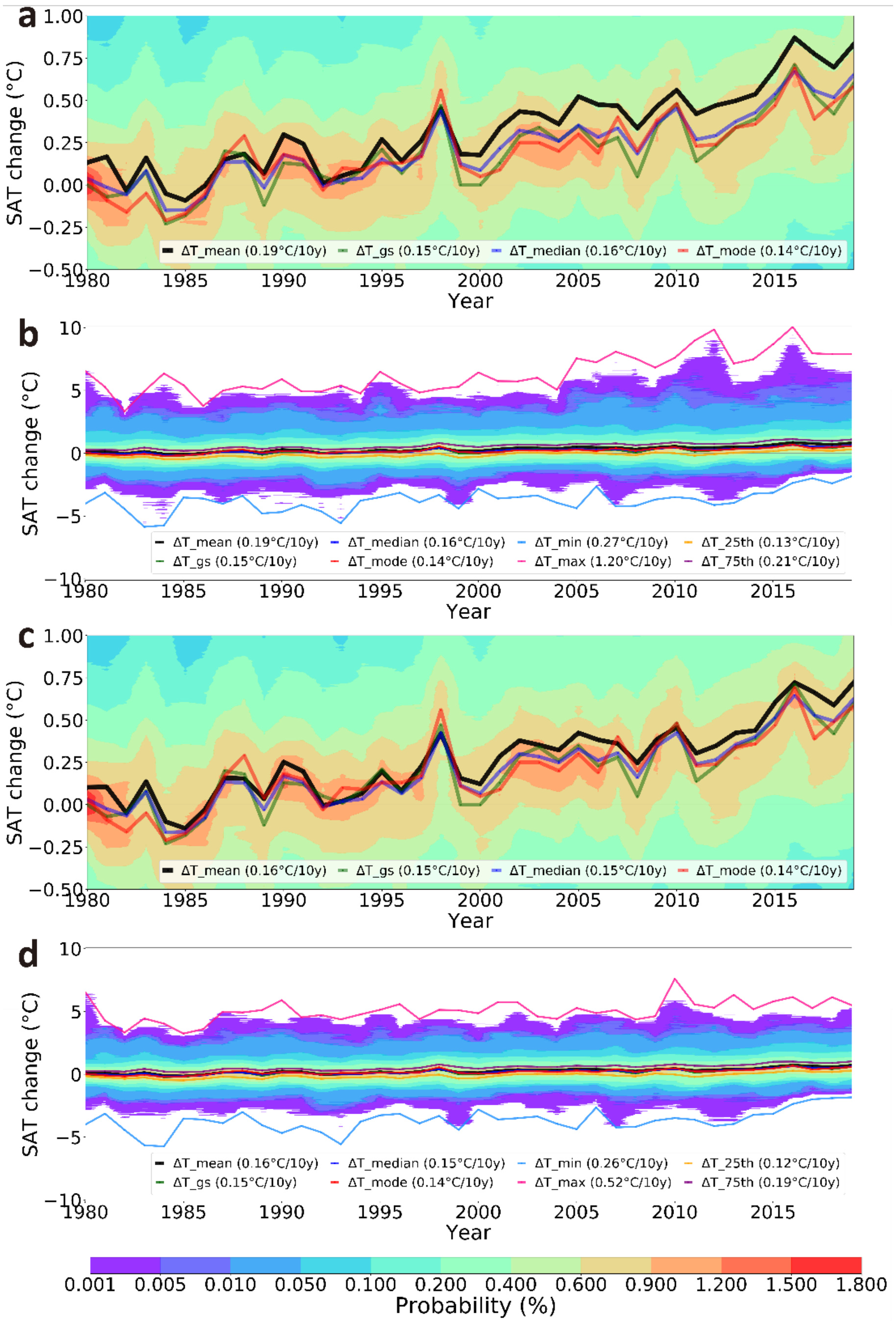

**Figure 2.** The influence of a SAT's asymmetrical nature on the value of global mean SAT change in the past four decades. (**a**) Time series (unit: °C) of $\Delta T_{mean}$ (black line), $\Delta T_{median}$ (blue line), $\Delta T_{mode}$ (red line) and $\Delta T_{gs}$ (green line) relative to the SAT in the base year of 1979. Shading indicates the image histogram value (unit: %) of SAT change of each year. (**b**) Same as (**a**) except plotted with a larger range of y-axis, and with time series (unit: °C) of $\Delta T_{min}$ (light blue line), $\Delta T_{max}$ (pink line), $\Delta T_{25th}$ (i.e. $25^{th}$ percentile of SAT change; yellow line) and $\Delta T_{75th}$ (i.e. $75^{th}$ percentile of SAT change; purple line) include. (**c**)–(**d**) Respectively the same as (**a**) and (**b**), except the Arctic region (65°N–90°N) is excluded from calculations. The orange and red shadings mark the SAT change values that are observed in a larger proportion of Earth surface. The linear trends, which are all statistically significant at the 99.9% confidence level, are marked in the figure legend. All results are obtained based on ERA5.

### *3.3 Impacts of asymmetrical SAT change on $\Delta T_{mean}$ in future climate projections*

In the famous Paris Agreement, 196 parties committed to maintaining the future global warming rate, in terms of $\Delta T_{mean}$, below 2°C compared to the preindustrial level to potentially mitigate severe climate impacts according to climate model simulations (Rogelj et al. 2015; Hof et al. 2017; Sanderson et al. 2017; Masson-Delmotte et al. 2018; Hoegh-Guldberg et al. 2019; Duan et al. 2021). We, based on twenty CMIP6 climate models (see section 2), find that the disagreement between $\Delta T_{mean}$ and $\Delta T_{gs}$ is even larger in climate model outputs that are used in current climate assessments and projections.

Among the historical runs (1979–2014) of twenty CMIP6 climate models, all produced larger increases in $\Delta T_{mean}$ than in $\Delta T_{median}$, $\Delta T_{mode}$ and $\Delta T_{gs}$. The ensemble long-term linear trend of $\Delta T_{mean}$ (0.24°C/decade) exceeds that of $\Delta T_{median}$ (0.19°C/decade), $\Delta T_{mode}$ (0.18°C/decade) and $\Delta T_{gs}$ (0.19°C/decade) by 0.05–0.06°C/decade (Fig. 3a). Although the increasing trends of all the four measures are overall larger than observation, their differences are consistent with the results obtained from modern reanalysis data. Similar to Fig. 2a, the ensemble $\Delta T_{mean}$ is generally above the SAT change values that are more frequently observed (red and orange shadings in Fig. 3a) over the world each year, as a result of the more positively skewed SAT change histogram and the faster increasing rate of large SAT change values (Figs. 3a and 3e). Consequently, the simulated $\Delta T_{mean}$ (0.52°C) from 1995–2014, relative to the SAT in 1979, exceeds $\Delta T_{gs}$ (0.42°C) by 0.10°C (Fig. 3i and Table 1). These results indicate that the ensemble mean of the twenty CMIP6 climate models are able to reproduce the proportion of the net warming effect of regional-scale temperature change to $\Delta T_{mean}$.

The ensemble results of the future climate projections of twenty models reflect further larger differences between $\Delta T_{mean}$ and global-scale warming. Under the emission scenario of SSP1-2.6, the increasing trend of $\Delta T_{mean}$ is 0.01–0.02°C/decade higher than those based on $\Delta T_{median}$, $\Delta T_{mode}$ and $\Delta T_{gs}$. In spite of the small trend differences, the difference between $\Delta T_{mean}$ and $\Delta T_{gs}$ reaches 0.27°C from 2080–2099 (Figs. 3b and 3i and Table 1). With fewer restrictions on greenhouse gas emissions, the $\Delta T_{mean}$ trends under the SSP2-4.5 and SSP5-8.5 scenarios are 0.03–0.04°C/decade and 0.06–0.07°C/decade larger than the $\Delta T_{median}$, $\Delta T_{mode}$ and $\Delta T_{gs}$ trends, respectively. $\Delta T_{mean}$ exceeds $\Delta T_{gs}$ by 0.35°C and 0.54°C from 2080–2099 under the two scenarios (Figs. 3c, 3d, 3i and Table 1). Obviously, the differences of $\Delta T_{mean}$ with $\Delta T_{gs}$, as well as with $\Delta T_{median}$ and $\Delta T_{mode}$, are proportional to the emission level. In either emission scenario, the $\Delta T_{median}$, $\Delta T_{mode}$ and $\Delta T_{gs}$ series all stay close to the SAT change values that are most frequently observed over the world (red shading in Figs. 3a–d), but the $\Delta T_{mean}$ series is exceptionally higher because the SAT change distribution becomes more asymmetrical under higher emission scenarios (Figs. 3f–h), implying the existence of more significant extreme regional-scale warming effect. In short, the difference between $\Delta T_{mean}$ and global-scale warming, that is the net regional-scale warming effect, becomes larger under higher emission scenarios because of the more asymmetrical SAT change distribution. Depending on the emission level, the predicted $\Delta T_{mean}$ could be 0.27–0.54°C greater than $\Delta T_{gs}$ by the end of the 21st century (Fig. 3e), which is a nonnegligible level compared to the 2°C target of the Paris Agreement.

As mentioned above, the difference between $\Delta T_{mean}$ and the global-scale warming rate is larger under scenarios with high emissions. Under high-emission scenarios (e.g., SSP2-4.5 and SSP5-8.5), the magnitudes of both skewness and variance of SAT histogram rapidly decrease (Fig. S4). This implies that the "business-as-usual" scenario tends to induce a more spatially heterogeneous warming effect, with faster warming in colder regions, and a greater change in the SAT histogram shape than do other scenarios, leading to a further increase in $\Delta T_{mean}$ (Fig. 3i). Namely, compared to the past observation, the difference between $\Delta T_{mean}$ and global-scale warming, as well as their ratio, will be different under different emission scenarios in future climate projection.

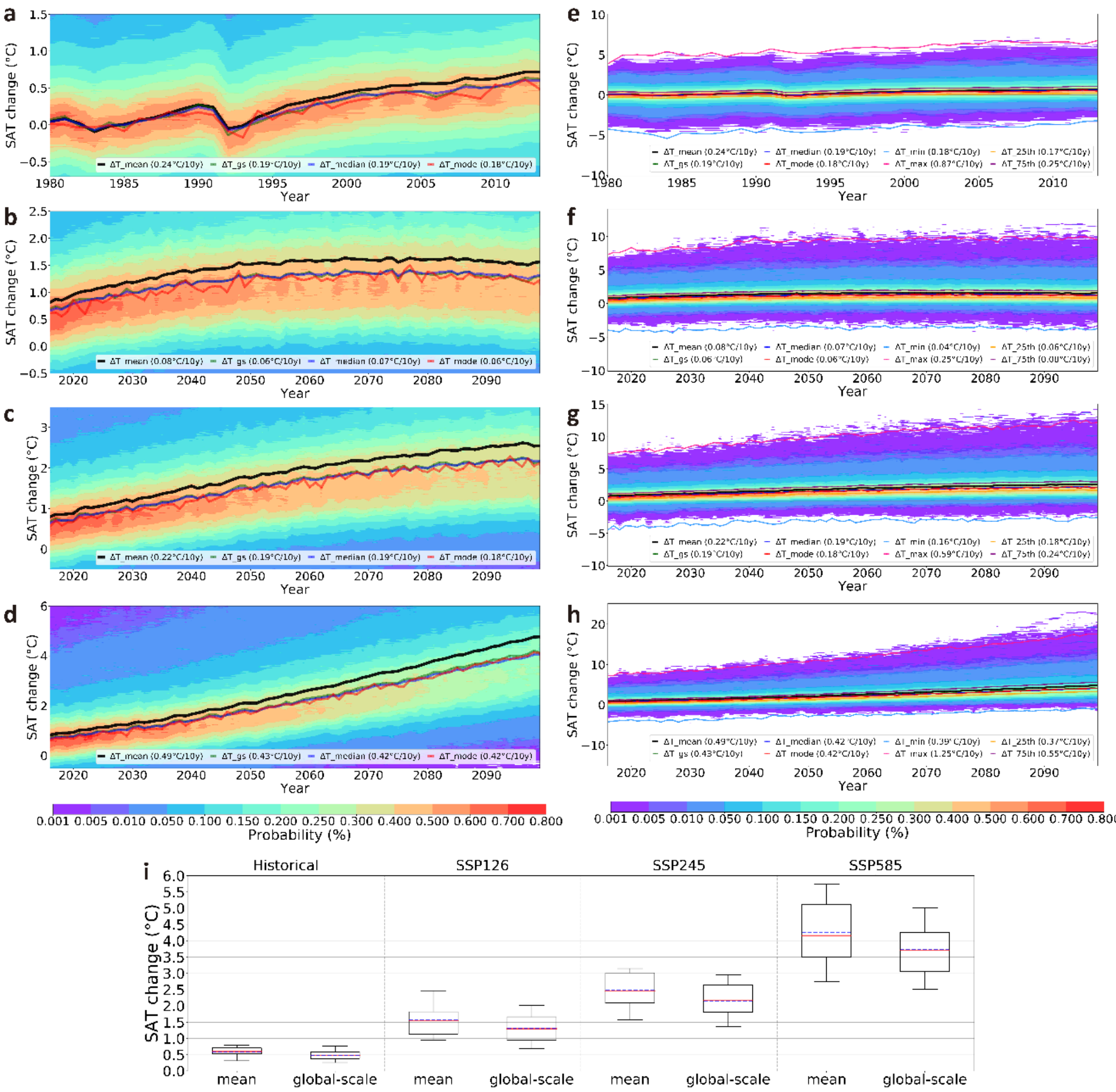

**Figure 3.** Excessively high Global mean SAT change in the CMIP6 model simulations. (**a**)–(**d**) Same as Fig. 2a, except for (**a**) the historical model run and the (**b**) SSP1-2.6, (**c**) SSP2-4.5, and (**d**) SSP5-8.5 projections based on the CMIP6 simulations. (**e**)–(**h**) Same as Fig. 2b, except for (**e**) the historical model run and the (**f**) SSP1-2.6, (**g**) SSP2-4.5, and (**h**) SSP5-8.5 projections based on the CMIP6 simulations. All panels are based on the ensemble results (unit: °C) of the twenty CMIP6 models. (**i**) Box plots of $\Delta T_{mean}$ (marked as "mean") and $\Delta T_{gs}$ (marked as "global-scale") from 2000–2014 for the historical run and 2080–2099 for future projections of climate model simulations relative to the SAT in the base year of 1979. The minimum, lower quartile, upper quartile, and maximum values are denoted by the boxes. The red solid line and blue dashed line in the boxplot indicate the median and mean values, respectively.

## 4. Discussion and conclusions

The analyses presented here indicate, based on the image histogram approach, that the global SAT changes are positively skewed and the asymmetrical nature of SAT change has significant impacts on the interpretation of global warming rate. The current definition of global warming rate, i.e. $\Delta T_{mean}$, is particularly susceptible to the influence of extreme regional-scale warming and produces a value larger than global-scale warming. Our results show that one could obtain very different global SAT changing rates based on different measures of central value (e.g. mean, median and mode), indicating a highly skewed SAT change distribution. Specifically, the global mean SAT increase from 2000–2019, relative to the SAT in 1979, is 0.19–0.23°C larger than the degree of global-scale warming. This implies the contribution of net positive regional-scale warming effect to $\Delta T_{mean}$ in recent decades, mainly due to exceptional warming in the Arctic region (i.e., Arctic amplification). By the end of the 21st century, the difference between $\Delta T_{mean}$ and $\Delta T_{gs}$ could reach 0.27–0.54°C under different emission scenarios according to CMIP6 model simulations, of which the projected global mean SAT increase is believed to be implausibly fast (Voosen 2019, 2021). In particular, the difference between $\Delta T_{mean}$ and $\Delta T_{gs}$, that is the net regional-scale warming effect, under the "business-as-usual" scenario (0.54°C) could be as high as 27% compared to the 2°C target of the Paris Agreement.

$\Delta T_{mean}$ is a consequence of both global-scale and regional-scale SAT changes, which could induce different climate and environmental impacts. However, regional-scale SAT changes are often larger in magnitude and have more direct impacts on society. For example, glacial retreat and sea-level rise are closely related to warming in polar regions, and climate impacts on ecosystems, biodiversity, agricultural production, and human health are strongly linked to regional-scale warming. Hence, these climate impacts may not be directly related to $\Delta T_{mean}$. This suggests that using $\Delta T_{mean}$ as the only measure of the global warming rate may not give accurate estimates of climate impacts if local- and regional-scale SAT changes are considerably different from the global-scale SAT trend, which are expected in high emission scenarios.

This paper constitutes a warning that extreme regional warming effects could have large impacts on the interpretation of $\Delta T_{mean}$, which is often considered in climate assessments and policies making, in both past observations and future climate projections. The above results also suggest the limitations and cautions inherent in using $\Delta T_{mean}$ as the only indicator of the global warming rate. While current climate policies, including mitigation and adaptation, are targeted at controlling the speed of global warming, an incorrect interpretation of the global warming rate could affect the policy decisions of policymakers and scientists' understanding of the mechanisms and impacts of climate change. Thus, on the way of seeking the best way to mitigate climate change, the limitations and shortcomings of the estimation of global warming rate should also be made very clear in related discussion, and more estimations of warming rate with different physical basis will help give a more comprehensive picture of global temperature change. In future climate assessments and policy decisions, the susceptibility of global mean SAT change to the asymmetrical warming distribution, as well as reasonable estimates of the global warming rate, should be considered.

## Data availability

Data relevant to the paper can be downloaded from the websites listed below:

- ERA5 provided by the Copernicus Climate Change Service Climate Data Store (CDS) at https://cds.climate.copernicus.eu/cdsapp#!/home;
- JRA55 provided by the Japan Meteorological Agency (JMA) at https://rda.ucar.edu/datasets/ds628.8/;
- NCEP2 provided by the NOAA/OAR/ESRL PSL at https://psl.noaa.gov/data/gridded/data.ncep.reanalysis2.surface.html;
- NCEP1 provided by the NOAA/OAR/ESRL PSL at https://psl.noaa.gov/data/gridded/data.ncep.reanalysis.html;
- CMIP6 model simulation at https://esgf-node.llnl.gov/projects/cmip6/.

## Competing interests

The authors declare no competing interests.

## Acknowledgements

The authors would like to thank Dr. Qingcun Zeng, Prof. Weihong Qian, Dr. Yuping Guan for their constructive comments on the earlier draft of this paper.

## Author contributions

J.C.H.L.: methodology, formal analysis, data curation, writing—original draf, writing—review and editing, visualization; B.Z.: conceptualization, supervision, methodology, writing—review and editing, funding acquisition; Y.G., J.X.L.W., W.Z., W.Q.: writing—review and editing. All authors reviewed the manuscript.

## Supplementary figures and tables

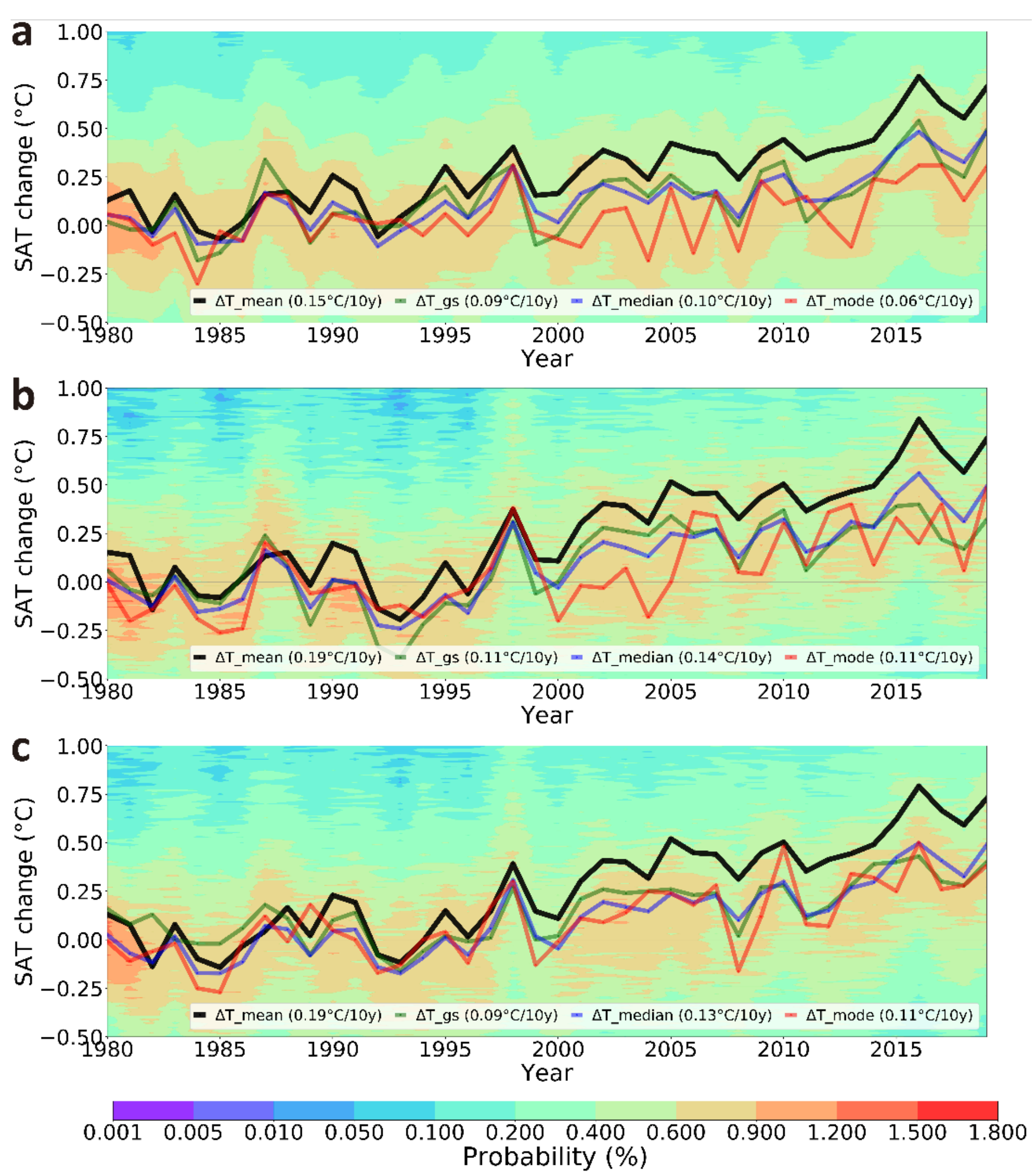

**Figure S1.** The influence of a SAT's asymmetrical nature on the value of global mean SAT change in the past four decades based on JRA55, NCEP1, and NCEP2. (**a**)–(**c**) Same as Fig. 2a, except based on (**a**) JRA55, (**b**) NCEP1, and (**c**) NCEP2. The linear trends, which are all significant at the 99.9% confidence level, are marked in the figure legend.

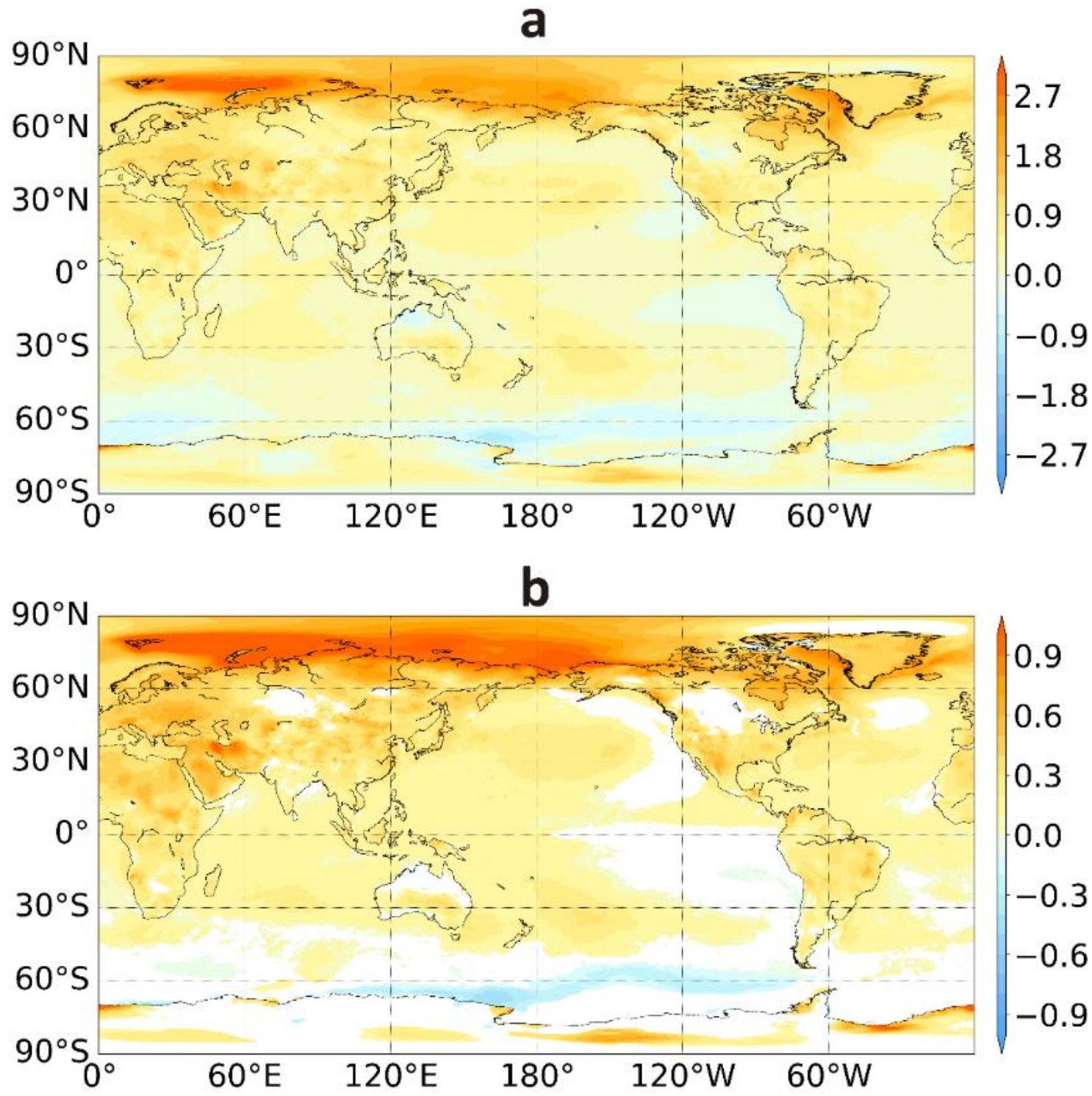

**Figure S2.** Spatial distribution of SAT change. (**a**)–(**b**) Spatial distributions of (**a**) the observed SAT difference (unit: °C) between 2000–2019 and 1980–1999 and that of (**b**) the observed linear SAT trend (unit: °C/decade) from 1979 to 2019. In (**b**), shading denotes significant linear trends (90% confidence level), and insignificant trends are not plotted (white).

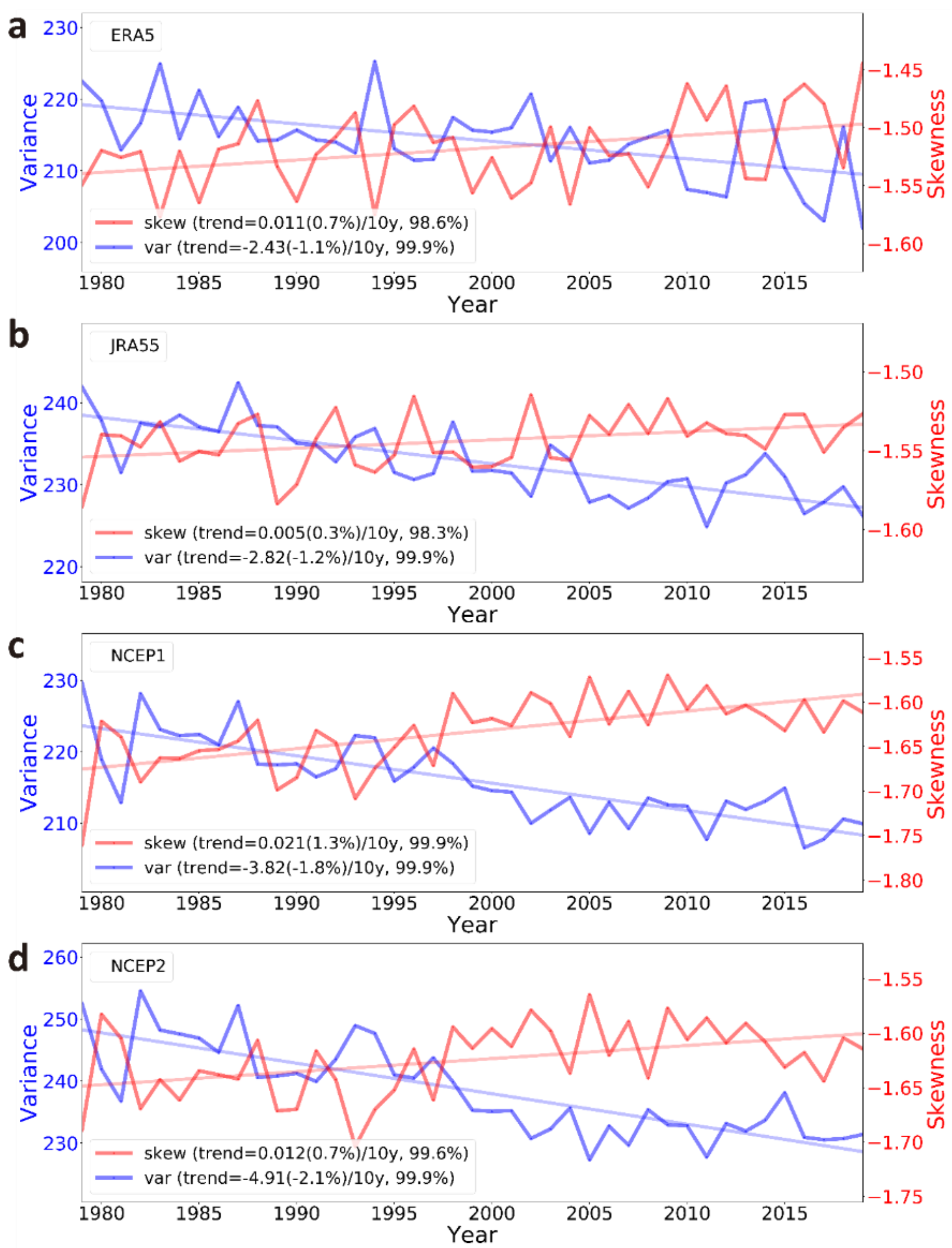

**Figure S3.** Decreases in the magnitudes of the variance and skewness of the SAT histogram from 1979 to 2019. (**a**)–(**d**) Time series of SAT variance (blue line, unit: °C$^2$) and skewness (red line, dimensionless) from 1979 to 2019 and the corresponding linear trends (thin lines) based on (**a**) ERA5, (**b**) JRA55, (**c**) NCEP1, and (d) NCEP2. All linear trends are statistically significant at the 99.9% confidence level, except those for SAT skewness in (**a**)–(**c**), which are statistically significant at the 98.6%, 98.3% and 99.6% level, respectively.

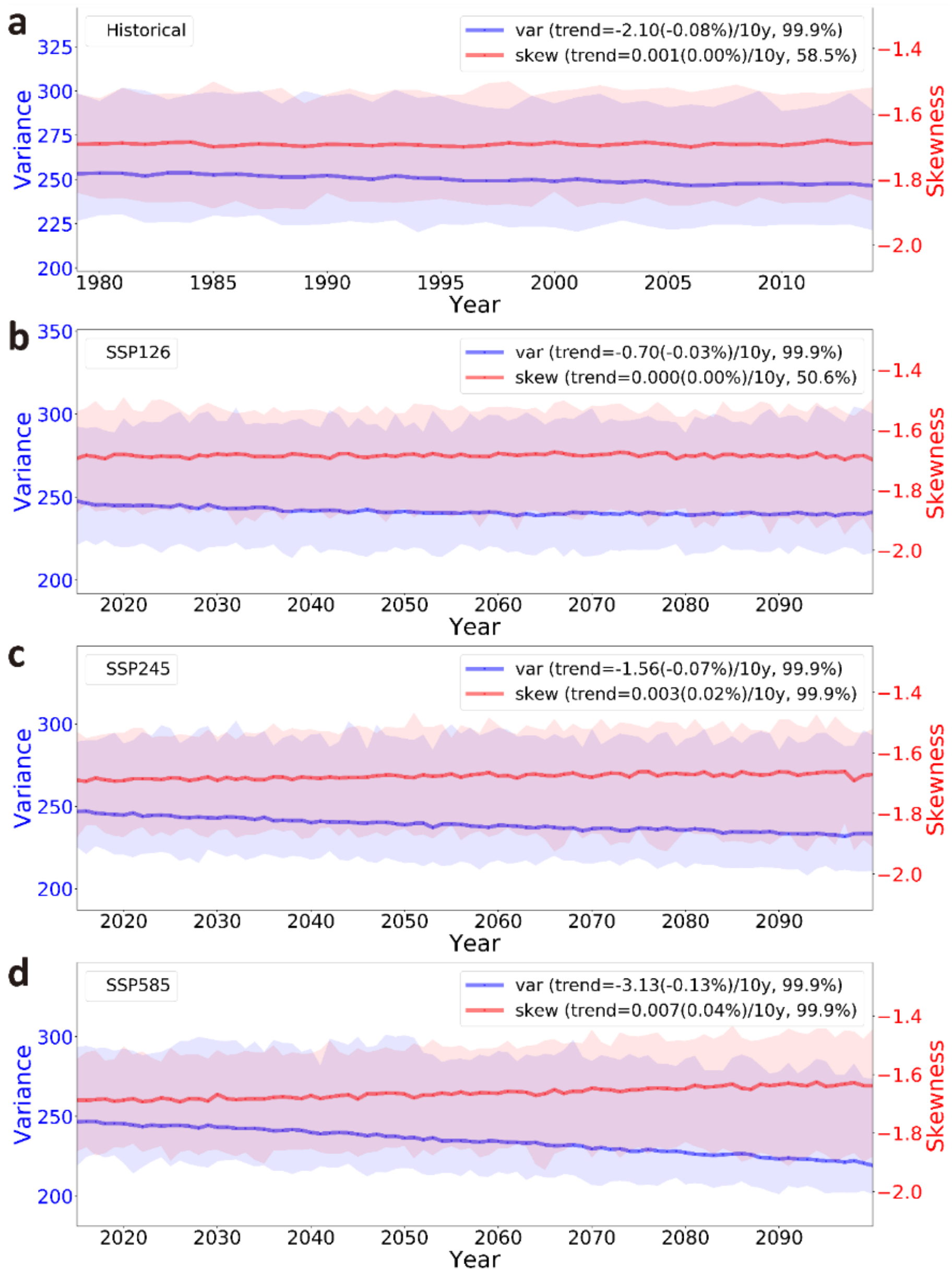

**Figure S4.** Changes in the variance and skewness of the SAT histogram for CMIP6 model simulations. (**a**)–(**d**) Same as Fig. S4, except for (**a**) model historical run and (**b**) SSP126, (**c**) SSP245, and (**d**) SSP585 projections. All panels are based on the ensemble results of the twenty CMIP6 models. The SAT histogram skewness and variance are denoted by "skew" and "var", respectively. Shading indicates the uncertainties of the model simulation results. All linear trends are statistically significant at the 99.9% confidence level, except those for SAT skewness in (**a**)–(**b**), which are statistically insignificant.